\documentclass[prl,twocolumn,overcite,superscriptaddress]{revtex4-1}
\usepackage[pdftex,colorlinks=true,allcolors=blue]{hyperref}
\usepackage{graphicx}
\usepackage{dcolumn}
\usepackage{bm}
\usepackage{color}
\usepackage{amsmath}
\usepackage{accents}

\begin{document}

\title{Nearly isotropic spin-pumping related Gilbert damping in Pt/Ni$_{81}$Fe$_{19}$/Pt}

\author{W. Cao}\email{wc2476@columbia.edu}
\affiliation{Materials Science and Engineering, Department of Applied Physics and Applied Mathematics, Columbia University, New York, New York 10027, USA}
\author{L. Yang}
\affiliation{Materials Science and Engineering, Department of Applied Physics and Applied Mathematics, Columbia University, New York, New York 10027, USA}
\author{S. Auffret}
\affiliation{SPINTEC, Universit$\acute{e}$ Grenoble Alpes/CEA/CNRS, F-38000 Grenoble, France}
\author{W.E. Bailey}\email{web54@columbia.edu}
\affiliation{Materials Science and Engineering, Department of Applied Physics and Applied Mathematics, Columbia University, New York, New York 10027, USA}
\affiliation{SPINTEC, Universit$\acute{e}$ Grenoble Alpes/CEA/CNRS, F-38000 Grenoble, France}

\date{\today}

\begin{abstract}
A recent theory by Chen and Zhang [Phys. Rev. Lett. 114, 126602 (2015)] predicts strongly anisotropic damping due to interfacial spin-orbit coupling in ultrathin magnetic films. Interfacial Gilbert-type relaxation, due to the spin pumping effect, is predicted to be significantly larger for magnetization oriented parallel to compared with perpendicular to the film plane. Here, we have measured the anisotropy in the Pt/Ni$_{81}$Fe$_{19}$/Pt system via variable-frequency, swept-field ferromagnetic resonance (FMR). We find a very small anisotropy of enhanced Gilbert damping with sign opposite to the prediction from the Rashba effect at the FM/Pt interface. The results are contrary to the predicted anisotropy and suggest that a mechanism separate from Rashba spin-orbit coupling causes the rapid onset of spin-current absorption in Pt.
\end{abstract}

\maketitle

\section{introduction}
The spin-transport properties of Pt have been studied intensively. Pt exhibits efficient, reciprocal conversion of charge to spin currents through the spin Hall effect (SHE)\cite{Saitoh_2006,Mosendz_2010,Liu_2011,Wang_2014}. It is typically used as detection layer for spin current evaluated in novel configurations\cite{Uchida_2008,Ellsworth_2016,Zhou_2016}. Even so, consensus has not yet been reached on the experimental parameters which characterize its spin transport. The spin Hall angle of Pt, the spin diffusion length of Pt, and the spin mixing conductance of Pt at different interfaces differ by as much as an order of magnitude when evaluated by different techniques\cite{Kurt_2002,Vila_2007,Ando_2008,Liu_2011,Mosendz_2010,Azevedo_2011,Althammer_2013}.

Recently, Chen and Zhang \cite{Chen_2015,Kai_Chen_2015} (hereafter CZ) have proposed that interfacial spin-orbit coupling (SOC) is a missing ingredient which can bring the measurements into greater agreement with each other. Measurements of spin-pumping-related damping, particularly, report spin diffusion lengths which are much shorter than those estimated through other techniques\cite{Feng_2012,Rojas_S_nchez_2014}. The introduction of Rashba SOC at the FM/Pt interface leads to interfacial spin-memory loss, with discontinuous loss of spin current incident to the FM/Pt interface. The model suggests that the small saturation length of damping enhancement reflects an interfacial discontinuity, while the inverse spin Hall effect (ISHE) measurements reflect the bulk absorption in the Pt layer\cite{Feng_2012,Rojas_S_nchez_2014}.

The CZ model predicts a strong anisotropy of the enhanced damping due to spin pumping, as measured in ferromagnetic resonance (FMR). The damping enhancement for time-averaged magnetization lying in the film plane ({\it pc}-FMR, or parallel condition) is predicted to be significantly larger than that for magnetization oriented normal to the film plane ({\it nc}-FMR, or normal condition). The predicted anisotropy can be as large as 30\%, with {\it pc}-FMR damping exceeding {\it nc}-FMR damping, as will be shown shortly.

In this paper, we have measured the anisotropy of the enhanced damping due to the addition of Pt in symmetric Pt/Ni$_{81}$Fe$_{19}$ (Py)/Pt structures. We find that the anisotropy is very weak, less than 5\%, and with the opposite sign from that predicted in \cite{Chen_2015}.

\section{theory}
We first quantify the CZ-model prediction for anisotropic damping due to the Rashba effect at the FM/Pt interface. In the theory, the spin-memory loss for spin current polarized perpendicular to the interfacial plane is always larger than that for spin current polarized in the interfacial plane. The pumped spin polarization $\bm{\sigma}=\bm{m} \times \dot{\bm{m}}$ is always perpendicular to the time-averaged or static magnetization $\langle\bm{m}\rangle_t \simeq \bm{m}$. For {\it nc}-FMR, the polarization $\bm{\sigma}$ of pumped spin current is always in the interfacial plane, but for {\it pc}-FMR, is nearly equally in-plane and out-of-plane. A greater damping enhancement is predicted in the {\it pc} condition than in the {\it nc} condition, $\Delta\alpha_{pc}>\Delta\alpha_{nc}$:

\begin{equation}
\Delta\alpha_{nc}=K\Big[\frac{1+4\eta\xi(t_{Pt})}{1+\xi(t_{Pt})}\Big]
\label{eqn1}
\end{equation}

\begin{equation}
\Delta\alpha_{pc}=K\Big[\frac{1+6\eta\xi(t_{Pt})}{1+\xi(t_{Pt})}+\frac{\eta}{2[1+\xi(t_{Pt})]^{2}}\Big]
\label{eqn2}
\end{equation}

\begin{equation}
\xi(t_{Pt})=\xi(\infty)\times\coth(t_{Pt}/\lambda_{sd})
\label{eqn3}
\end{equation}

where the constant of proportionality K is the same for both conditions and the dimensionless parameters, $\eta$ and $\xi$, are always real and positive. The Rashba parameter

\begin{equation}
\eta=(\alpha_{R}k_{F}/E_{F})^{2}
\label{eqn4}
\end{equation}

is proportional to the square of the Rashba coefficient $\alpha_{R}$, defined as the strength of the Rashba potential, $V(\bm{r})=\alpha_{R}\delta(z)(\boldsymbol{\hat{k}}\times\boldsymbol{\hat{z}})\cdot\bm{\sigma}$, where $\delta(z)$ is a delta function localizing the effect to the interface at $z=0$ (film plane is {\it xy}), $k_{F}$ is the Fermi wavenumber, and $E_{F}$ is the Fermi energy. The backflow factor $\xi$ is a function of Pt layer thickness, where the backflow fraction at infinitely large Pt thickness defined as $\epsilon=\xi(\infty)/[1+\xi(\infty)]$. $\epsilon=\textrm{0 (1)}$ refers to zero (complete) backflow of spin current across the interface. $\lambda_{sd}$ is the spin diffusion length in the Pt layer.

\begin{figure}
  \includegraphics[width=\columnwidth]{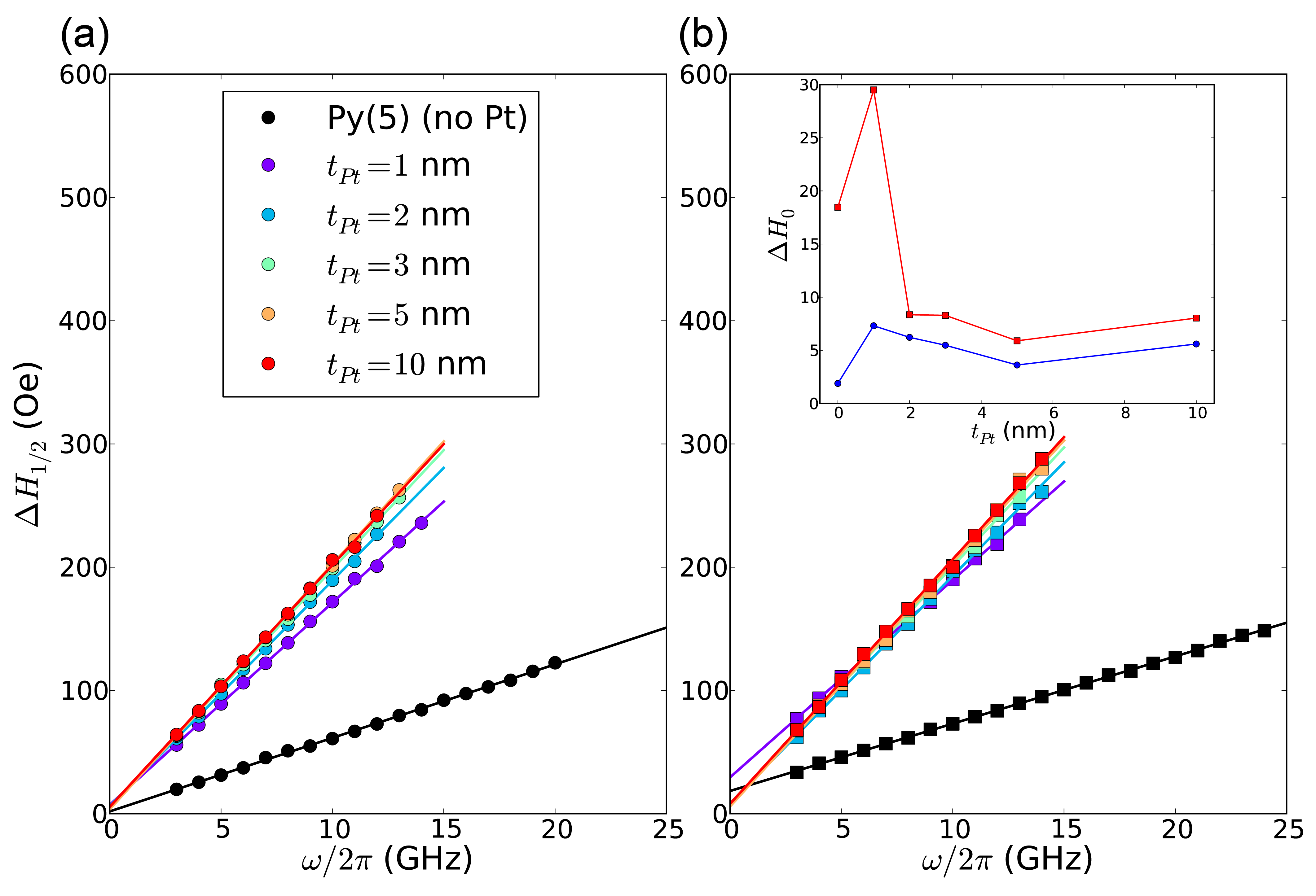}
  \caption{Frequency-dependent half-power FMR linewidth $\Delta H_{1/2}(\omega)$ of the reference sample Py(5 nm) (black) and symmetric trilayer samples Pt(t)/Py(5 nm)/Pt(t) (colored). (a) {\it pc}-FMR measurements. (b) {\it nc}-FMR measurements. Solid lines are linear fits to extract Gilbert damping $\alpha$. (Inset): inhomogeneous broadening $\Delta H_0$ in {\it pc}-FMR (blue) and {\it nc}-FMR (red).}
  \label{fig1}
\end{figure}

To quantify the anisotropy of the damping, we define Q:

\begin{equation}
Q\equiv(\Delta\alpha_{pc}-\Delta\alpha_{nc})/\Delta\alpha_{nc}
\label{eqn5}
\end{equation}

as an {\it anisotropy factor}, the fractional difference between the enhanced damping in pc and nc conditions. Positive Q (Q\textgreater0) is predicted by the CZ model. A spin-memory loss $\delta$ factor of 0.9 $\pm$ 0.1, corresponding to nearly complete relaxation of spin current at the interface with Pt, was measured through current perpendicular to plane-magnetoresistance (CPP-GMR)\cite{Kurt_2002} According to the theory\cite{Chen_2015,Kai_Chen_2015}, the spin-memory loss can be related to the Rashba parameter by $\delta=2\eta$, so we take $\eta\sim0.45$. The effect of variable $\eta<0.45$ will be shown in Figure \ref{fig3}. To evaluate the thickness dependent backflow $\xi(t_{Pt})$, we assume $\lambda_{sd}^{Pt}=14$ nm, which is associated with the absorption of the spin current in the bulk of Pt layer, as found from CPP-GMR measurements\cite{Kurt_2002} and cited in \cite{Chen_2015}. Note that this $\lambda_{sd}^{Pt}$ is longer than that used sometimes to fit FMR data\cite{Feng_2012,Rojas_S_nchez_2014}; Rashba interfacial coupling in the CZ model brings the onset thickness down. The calculated anisotropy factor Q should then be as large as 0.3, indicating that $\Delta\alpha_{pc}$ is 30\% greater than $\Delta\alpha_{nc}$ (see Results for details).

\begin{figure}
  \includegraphics[width=\columnwidth]{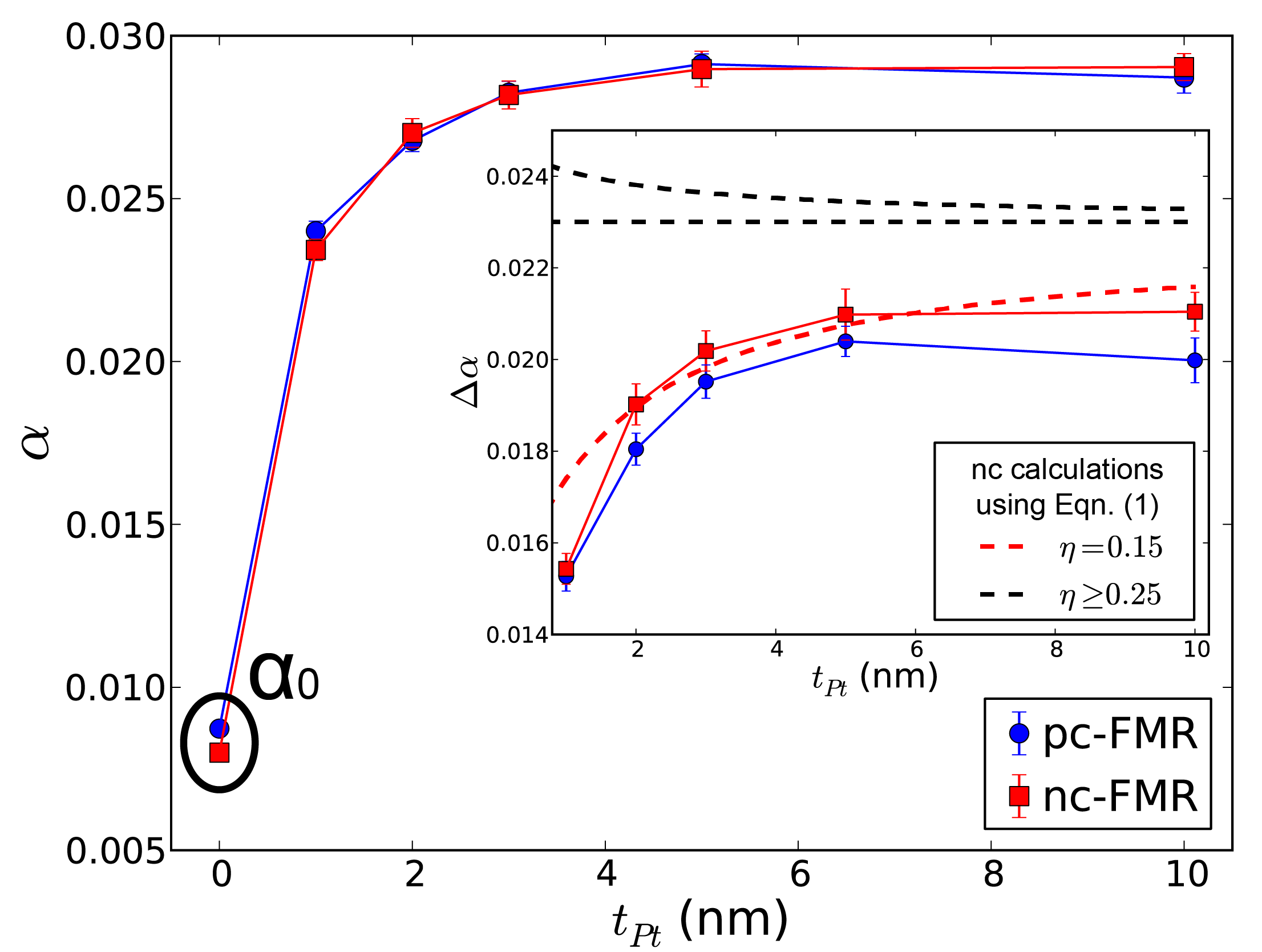}
  \caption{Pt thickness dependence of Gilbert damping $\alpha=\alpha(t_{Pt})$ in {\it pc}-FMR (blue) and {\it nc}-FMR (red). $\alpha_{0}$ refers to the reference sample ($t_{Pt}=0$). (Inset): Damping enhancement $\Delta \alpha(t_{Pt})=\alpha(t_{Pt})-\alpha_{0}$ due to the addition of Pt layers in {\it pc}-FMR (blue) and {\it nc}-FMR (red). Dashed lines refer to calculated $\Delta \alpha_{nc}$ using Equation \ref{eqn1} by assuming $\lambda_{sd}^{Pt}=14$ nm and $\epsilon=10\%$. The red dashed line ($\eta=0.15$) shows a similar curvature with experiments; The black dashed line ($\eta \geq 0.25$) shows a curvature with the opposite sign.}
  \label{fig2}
\end{figure}

\begin{figure*}
  \includegraphics[width=\textwidth]{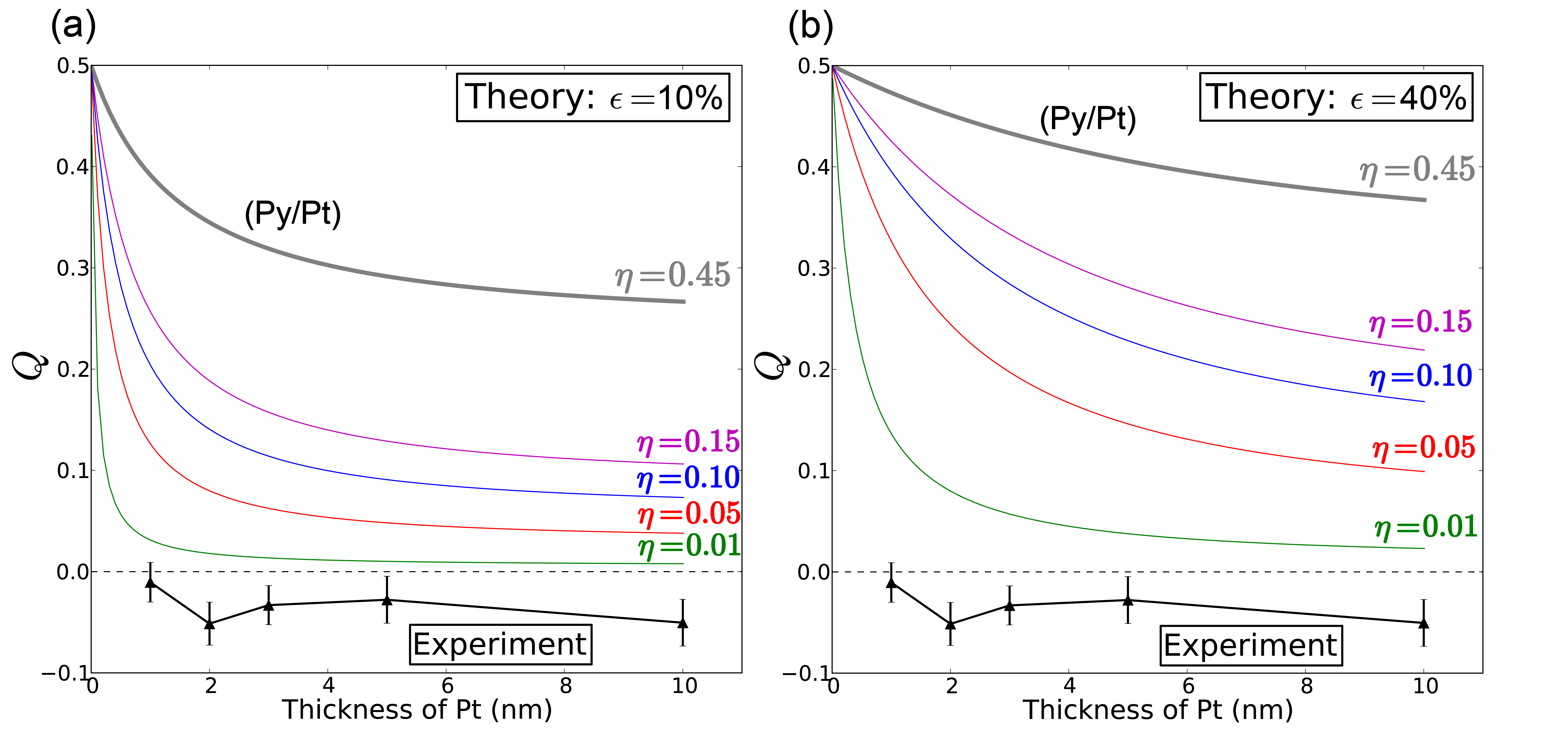}
  \caption{Anisotropy factor Q for spin-pumping enhanced damping, defined in Equation \ref{eqn5}. Solid lines are calculations using the CZ theory\cite{Chen_2015}, Equations \ref{eqn1}--\ref{eqn3}, for variable Rashba parameter $0.01\leq\eta\leq0.45$. $\lambda_{sd}^{Pt}$ is set to be 14 nm. Backflow fraction $\epsilon$ is set to be 10\% in (a) and 40\% in (b). Black triangles, duplicate in (a) and (b), show the experimental values from Figure 2.}
  \label{fig3}
\end{figure*}

\section{experiment}
In this paper, we present measurements of the anisotropy of damping in the symmetric Pt($t_{Pt}$)/Py(5 nm)/Pt($t_{Pt}$) system, where ``Py''=Ni$_{81}$Fe$_{19}$. Because the Py thickness is much thicker than its spin coherence length\cite{GhostPRL}, we expect that spin-pumping-related damping at the two Py/Pt interfaces will sum. The full deposited stack is Ta(5 nm)/Cu(5 nm)/Pt($t_{Pt}$)/Py(5 nm)/Pt($t_{Pt}$)/Al$_2$O$_3$(3 nm), $t_{Pt}=\textrm{1--10 nm}$, deposited via DC magnetron sputtering under computer control on ion-cleaned Si/SiO$_{2}$ substrates at ambient temperature. The deposition rates were 0.14 nm/s for Py and 0.07 nm/s for Pt. Heterostructures deposited identically, in the same deposition chamber, have been shown to exhibit both robust spin pumping effects, as measured through FMR linewidth\cite{Ghosh2011,Caminale_2016}, and robust Rashba effects (in Co/Pt), as measured through Kerr microscopy\cite{MironNM2010,MironNM2011}. The stack without Pt layers was also deposited as the reference sample. The films were characterized using variable frequency FMR on a coplanar waveguide (CPW) with center conductor width of 300 $\mu$m. The bias magnetic field was applied both in the film plane ({\it pc}) and perpendicular to the plane ({\it nc}), as previously shown in \cite{Yang_2016}. The {\it nc}-FMR measurements require precise alignment of the field with respect to the film normal. Here, samples were aligned by rotation on two axes to maximize the resonance field at 3 GHz.

\section{results and analysis}
Figure \ref{fig1} shows frequency-dependent half-power linewidth $\Delta H_{1/2}(\omega)$ in {\it pc}- and {\it nc}-FMR. The measurements were taken at frequencies from 3 GHz to a cut-off frequency above which the signal-to-noise ratio becomes too small for reliable measurement of linewidth. The cutoff ranged from 12--14 GHz for the samples with Pt (linewidth $\sim$ 200--300 G) to above 20 GHz for $t_{Pt}=0$. Solid lines stand for linear regression of the variable-frequency FMR linewidth $\Delta H_{1/2}=\Delta H_{0}+2\alpha\omega/\gamma$, where $\Delta H_{1/2}$ is the full-width at half-maximum, $\Delta H_{0}$ is the inhomogeneous broadening, $\alpha$ is the Gilbert damping, $\omega$ is the resonance frequency and $\gamma$ is the gyromagnetic ratio. The fits show good linearity with frequency $\omega/2\pi$ for all experimental linewidths $\Delta H_{1/2}(\omega)$. The inset summarizes inhomogeneous broadening $\Delta H_{0}$ in {\it pc}- and {\it nc}-FMR; its errorbar is $\sim 2$ Oe.

In Figure \ref{fig2}, we plot Pt thickness dependence of damping parameters $\alpha(t_{Pt})$ extracted from the linear fits in Figure \ref{fig1}, for both {\it pc}-FMR and {\it nc}-FMR measurements. Standard deviation errors in the fits for $\alpha$ are $\sim 3\times10^{-4}$. The Gilbert damping $\alpha$ saturates quickly as a function of $t_{Pt}$ in both pc and nc conditions, with 90\% of the effect realized with Pt(3 nm). The inset shows the damping enhancement $\Delta\alpha$ due to the addition of Pt layers $\Delta\alpha=\alpha-\alpha_{0}$, normalized to the Gilbert damping $\alpha_{0}$ of the reference sample without Pt layers. The Pt thickness dependence of $\Delta\alpha$ matches our previous study on Py/Pt heterostructures\cite{Caminale_2016} reasonably; the saturation value of $\Delta\alpha_{Pt/Py/Pt}$ is 1.7x larger than that measured for the single interface $\Delta\alpha_{Py/Pt}$\cite{Caminale_2016} (2x expected). The dashed lines in the inset refer to calculated $\Delta \alpha_{nc}$ using Equation \ref{eqn1} (assuming $\lambda_{sd}^{Pt}=14$ nm and $\epsilon=10\%$). $\eta=0.25$ shows a threshold of Pt thickness dependence. When $\eta>0.25$, the curvature of $\Delta \alpha(t_{Pt})$ will have the opposite sign to that observed in experiments, so $\eta=0.25$ is the maximum which can qualitatively reproduce the Pt thickness dependence of the damping.

As shown in Figure \ref{fig2} inset, the damping enhancement due to the addition of Pt layers is slightly larger in the {\it nc} geometry than in the {\it pc} geometry: $\Delta\alpha_{nc}>\Delta\alpha_{pc}$. This is opposite to the prediction of the model in \cite{Chen_2015}. The anisotropy factor $Q\equiv(\Delta\alpha_{pc}-\Delta\alpha_{nc})/\Delta\alpha_{nc}$ for the model (Q\textgreater0) and the experiment (Q\textless0) are shown together in Figure \ref{fig3} (a) and (b). The magnitude of Q for the experiment is also quite small, with -0.05\textless Q\textless0. This very weak anisotropy, or near isotropy, of the spin-pumping damping is contrary to the prediction in \cite{Chen_2015}, and is the central result of our paper.

The two panels (a) and (b), which present the same experimental data (triangles), consider different model parameters, corresponding to negligible backflow ($\epsilon=0.1$, panel {\it a}) and moderate backflow ($\epsilon=0.4$, panel {\it b}) for a range of Rashba couplings $0.01\le \eta \le0.45$. A spin diffusion length $\lambda_{sd}=14$ nm for Pt\cite{Kurt_2002} was assumed in all cases.

The choice of backflow fraction $\epsilon=0.1$ or $0.4$ and the choice of spin diffusion length of Pt $\lambda_{sd}=14$ nm follow the CZ paper\cite{Chen_2015} for better evaluation of their theory. For good spin sinks like Pt, the backflow fraction is usually quite small. If $\epsilon=0$, then there will be no spin backflow. In this limit, $\Delta \alpha_{pc}$, $\Delta \alpha_{nc}$ and the Q factor will be independent of Pt thickness.

In the case of a short spin diffusion length of Pt, e.g., $\lambda_{sd}=3$ nm, the anisotropy Q as a function of Pt thickness decreases more quickly for ultrathin Pt, closer to our experimental observations. However, we note that the CZ theory requires a long spin diffusion length in order to reconcile different experiments, particularly CPP-GMR with spin pumping, and is not relevant to evaluate the theory in this limit.

Leaving apart the question of the sign of Q, we can see that the observed absolute magnitude is lower than that predicted for $\eta=0.05$ for small backflow and 0.01 for moderate backflow. According to ref \cite{Chen_2015}, a minimum level for the theory to describe the system with strong interfacial SOC is $\eta=0.3$.

\section{discussion}
Here, we discuss extrinsic effects which may result in a discrepancy between the CZ model (Q$\sim$+0.3) and our experimental result (-0.05\textless Q\textless0). A possible role of two-magnon scattering\cite{Arias1999,McMichael2004}, known to be an anisotropic contribution to linewidth $\Delta H_{1/2}$, must be considered. Two-magnon scattering is present for {\it pc}-FMR and nearly absent for {\it nc}-FMR. This mechanism does not seem to play an important role in the results presented. It is difficult to locate a two-magnon scattering contribution to linewidth in the pure Py film: Figure \ref{fig1} shows highly linear $\Delta H_{1/2}(\omega)$, without offset, over the full range to $\omega /2\pi=20$ GHz, thereby reflecting Gilbert-type damping. The damping for this film is much smaller than that added by the Pt layers. If the introduction of Pt adds some two-magnon linewidth, eventually mistaken for intrinsic Gilbert damping $\alpha$, this could only produce a measurement of Q\textgreater0, which was not observed.

One may also ask whether the samples are appropriate to test the theory. The first question regards sample quality. The Rashba Hamiltonian models a very abrupt interface. Samples deposited identically, in the same deposition chamber, have exhibited strong Rashba effects, so we expect the samples to be generally appropriate in terms of quality. Intermixing of Pt in Ni$_{81}$Fe$_{19}$ (Py)/Pt\cite{Golod2011} may play a greater role than it does in Co/Pt\cite{BERTERO1994173}, although defocused TEM images have shown fairly well-defined interfaces for our samples\cite{Bailey2012}.

A second question might be about the magnitude of the Rashba parameter $\eta$ in the materials systems of interest. Our observation of nearly isotropic damping is consistent with the theory, within experimental error and apart from the opposite sign, if the Rashba parameter $\eta$ is very low and the backflow fraction $\epsilon$ is very low. Ab-initio calculations for (epitaxial) Co/Pt in the ref\cite{Grytsyuk2016} have indicated $\eta=\textrm{0.02--0.03}$, lower than the values of $\eta \sim$ 0.45 assumed in \cite{Chen_2015,Kai_Chen_2015} to treat interfacial spin-memory loss.

The origin of the small, negative Q observed here is unclear. A recent paper has reported that $\Delta \alpha_{pc}$ is smaller than $\Delta \alpha_{nc}$ in the YIG/Pt system via single-frequency, variable-angle measurements\cite{Zhou_2016}, which is contrary to the CZ model prediction as well. It is also possible that a few monolayers of Pt next to the Py/Pt interfaces are magnetized in the samples\cite{Caminale_2016}, and this may have an unknown effect on the sign, not taken into account in the theory.

\section{conclusions}
In summary, we have experimentally demonstrated that in Pt/Py/Pt trilayers the interfacial damping attributed to spin pumping is nearly isotropic, with an anisotropy between film-parallel and film-normal measurements of \textless5\%. The nearly isotropic character of the effect is more compatible with conventional descriptions of spin pumping than with the Rashba spin-memory loss model predicted in \cite{Chen_2015}.

\section{acknowledgements}
We acknowledge support from the US NSF-DMR-1411160 and the Nanosciences Foundation, Grenoble.

\bibliography{ref}

\end{document}